\documentclass[useAMS,usenatbib]{mn2e}

\usepackage{graphicx}
\usepackage{hyperref}
\usepackage{upgreek}
\usepackage{multirow}
\usepackage{longtable}

\newcommand{\arxiv}[1]{(\href{http://arxiv.org/abs/#1}{arXiv:#1})}

\newcommand{\arxivpreprint}[1]{preprint (\href{http://arxiv.org/abs/#1}{arXiv:#1})}

\newcommand{\changed}[1]{#1}

\newcommand{\object}[1]{#1}

\def\aj{AJ}					
\def\apj{ApJ}					
\def\mnras{MNRAS}					
\def\apjs{ApJS}					
\def\aap{A\&A}					
\def\aaps{A\&AS}				

\title{One Centimetre Receiver Array-prototype observations of the CRATES sources at 30 GHz}
\author[M.\,W. Peel et al.]
 {M. W. Peel$^1$, M.\,P. Gawro\'nski$^2$, R.\,A.~Battye$^1$, M.~Birkinshaw$^3$, I.\,W.\,A.~Browne$^1$,
 \newauthor R.\,J.~Davis$^1$, R.~Feiler$^2$, A.\,J.~Kus$^2$, K.~Lancaster$^3$, S.\,R.~Lowe$^1$, B.\,M.~Pazderska$^2$,
 \newauthor E.~Pazderski$^2$, B.\,F.~Roukema$^2$ and P.\,N.~Wilkinson$^1$\\
 $^1$ Jodrell Bank Centre for Astrophysics, The University of Manchester, Manchester, M13 9PL\\
 $^2$ Toru\'n Centre for Astronomy, Nicolaus Copernicus University, 87-100 Toru\'n/Piwnice, Poland \\
 $^3$ University of Bristol, Tyndall Avenue, Bristol, BS8 ITL
}
 
\begin{document}

\date{Accepted 2010 September 4. Received 2010 September 1; in original form 2010 July 29}

\pagerange{\pageref{firstpage}--\pageref{lastpage}} \pubyear{2010}

\maketitle

\label{firstpage}

\begin{abstract}
Knowledge of the population of radio sources in the range $\sim2-200$~GHz is important for understanding their effects on measurements of the Cosmic Microwave Background power spectrum. We report measurements of the 30~GHz flux densities of 605 radio sources from the Combined Radio All-sky Targeted Eight-GHz Survey (CRATES), which have been made with the One Centimetre Receiver Array prototype (OCRA-p) on the Toru\'n 32-m telescope. The flux densities of sources that were also observed by WMAP and previous OCRA surveys are in broad agreement with those reported here, however a number of sources display intrinsic variability. We find a good correlation between the 30~GHz and Fermi gamma-ray flux densities for common sources. We examine the radio spectra of all observed sources and report a number of Gigahertz-peaked and inverted spectrum sources. These measurements will be useful for comparison to those from the Low Frequency Instrument of the \it Planck \rm satellite, which will make some of its most sensitive observations in the region covered here.
\end{abstract}


\maketitle

\section{Introduction}
Emission from blazars dominates the high latitude sky at high radio frequencies and also at gamma-ray frequencies. Blazars are a major foreground contaminant of observations of the Cosmic Microwave Background (CMB), particularly at high multipoles. In order to subtract effects of such objects from CMB observations, it is necessary to know the flux densities of individual bright sources as well as the statistical properties of the overall source population. Knowledge of the brightest individual sources comes directly from the CMB surveys themselves. WMAP has detected sources down to $\sim$~0.5~Jy at 22-93~GHz \citep{2010Gold} and the more sensitive Low Frequency Instrument (LFI) in the {\it Planck} satellite will detect sources with flux densities of a few hundred mJy at 33, 44 and 70~GHz. However, a knowledge of the statistical properties of significantly weaker sources is desirable.

At the present time there are no point source surveys with the appropriate combinations of flux density limit and frequencies to understand the contaminating effect that these sources have on CMB experiments such as {\it Planck}. Thus at present one must rely on measuring the high frequency properties of sources selected from lower frequency surveys in order to infer the high frequency population statistics. We have a programme aimed at characterising the high frequency radio source population in total intensity (\citealp{2007Lowe}; \citealp{2009Gawronski}) and polarisation (\citealp{2009Jackson}; \citealp{2010Battye}). The present paper is the latest in this series, being intermediate in flux density between the strong ($>$350~mJy at 4.85~GHz) Caltech-Jodrell Bank flat-spectrum (CJF) sample \citep{2007Lowe} and the weaker Very Small Array (VSA) sources \citep{2009Gawronski}.

\begin{figure*}
\centering
\includegraphics[scale=0.62]{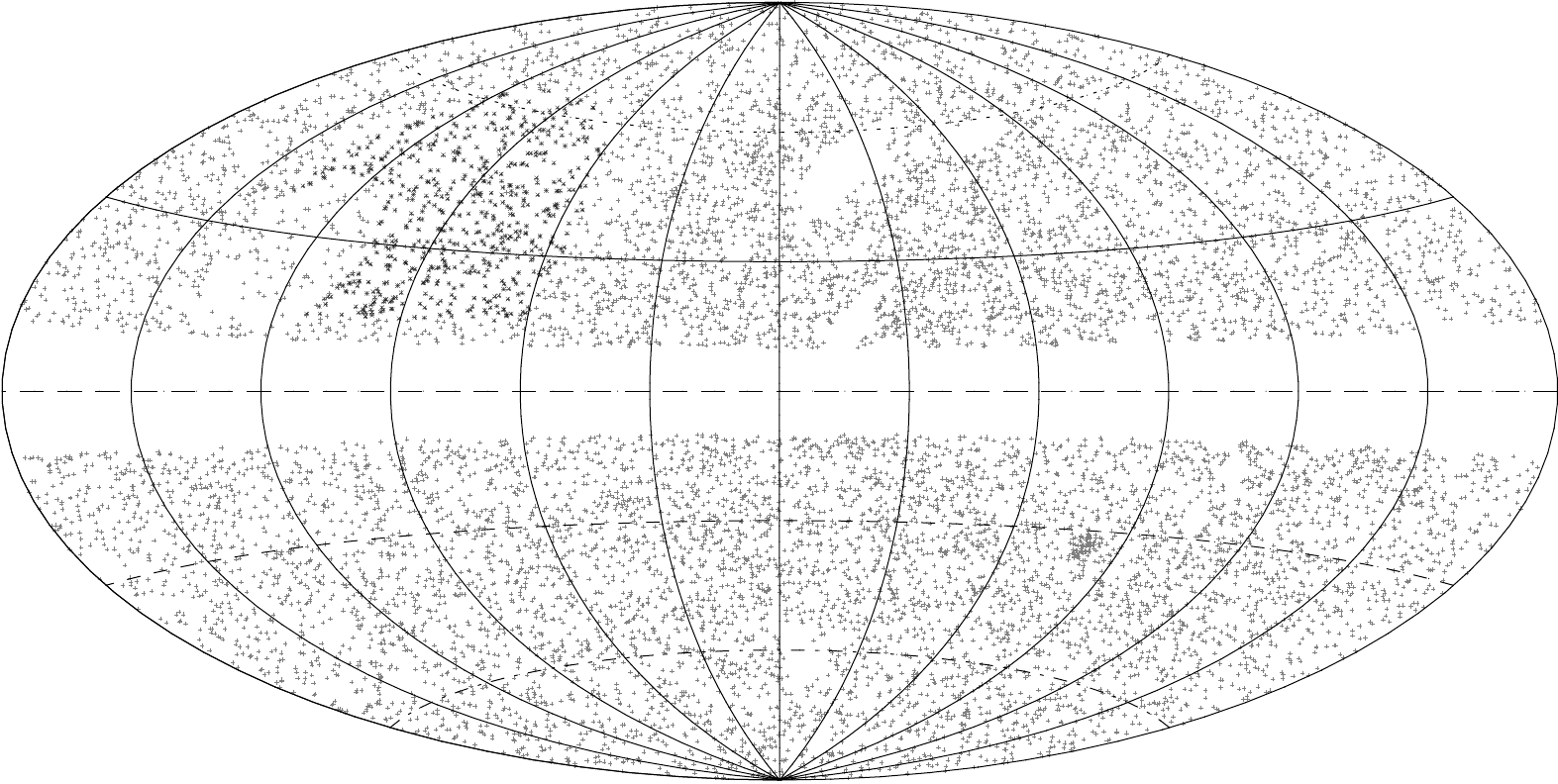}
\includegraphics[scale=0.58]{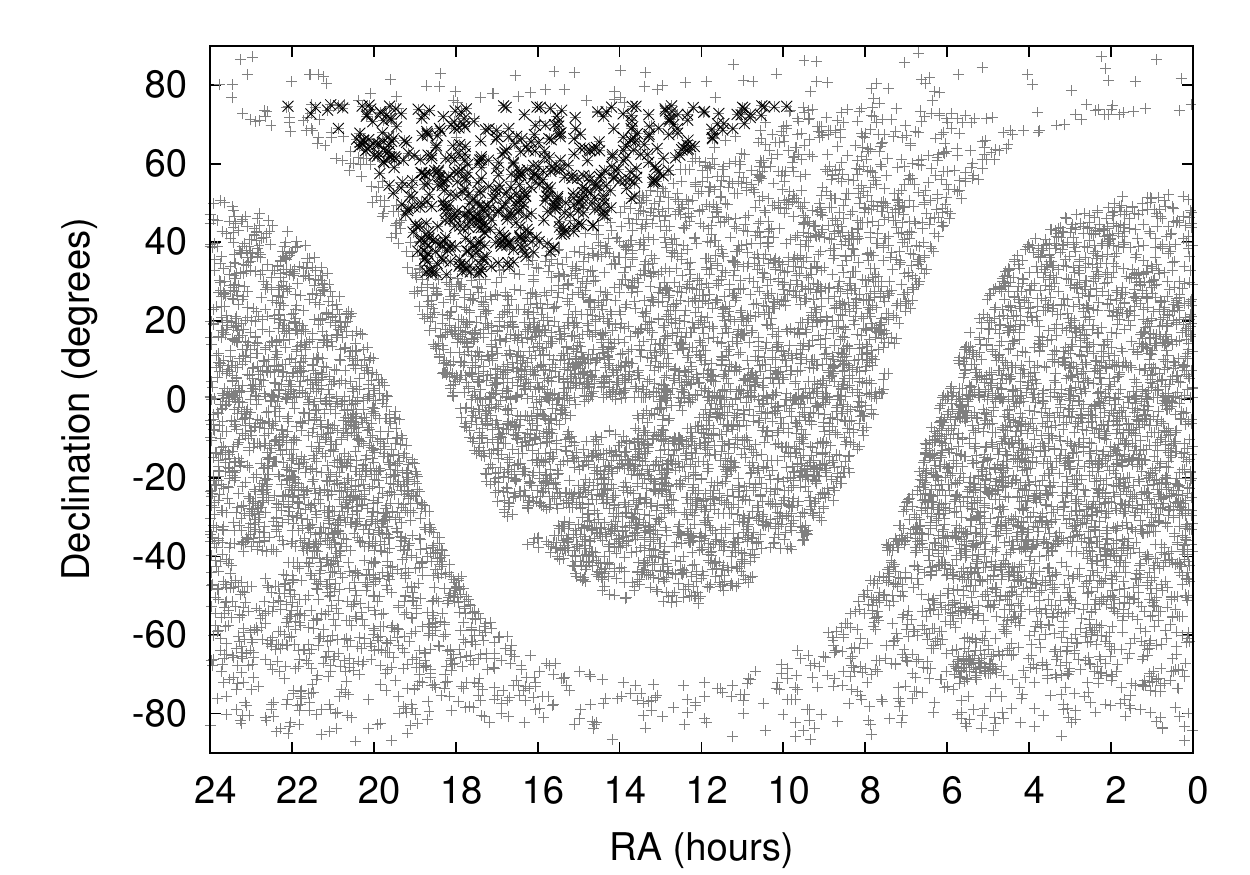}
\caption{\changed{The distribution of CRATES sources in Galactic coordinates in Aitoff equal-area projection centred on $l=0^\circ$ (left) and in RA and Dec (right). The complete CRATES sample is shown in grey. The sample observed by OCRA is shown in black. The small cluster of sources at low declination is the Large Magellanic Cloud, and the empty areas below Dec 0$\ensuremath{^\circ}$ are where there are gaps in the 4.8~GHz observation coverage \citep{2007Healey}. The reduced number of sources above 75$\ensuremath{^\circ}$ is caused by the upper declination limit in the GB6 observations.}}
\label{fig:crates_sources_aitoff}
\end{figure*}

The Combined Radio All-sky Targeted Eight GHz Survey (CRATES; \citealp{2007Healey}) is a sample of $\sim$11~000 strong flat spectrum sources with measured flux densities at 8.4~GHz. CRATES is currently the most complete large-area, flat spectrum point source sample at flux densities of hundreds of mJy. It samples a flux density range starting over an order of magnitude lower than the WMAP source sample \citep{2010Gold}. Thus CRATES sources represent excellent targets to follow up at higher frequencies.

The CRATES sample was originally selected to study blazars -- radio-loud Active Galactic Nuclei in which the relativistic jet axis points close to the observer's line of sight. This angle of observation means that the emission is Doppler boosted such that the received flux densities are dramatically increased compared to what would be observed at larger angles to the jet axis. This beamed non-thermal emission stretches from radio to gamma-ray frequencies. Blazars dominate the high Galactic latitude gamma-ray source counts produced by the Fermi Gamma-ray Space Telescope \citep{2010Abdoa,2010Abdob} and many will be detected by Planck, SCUBA2 and Herschel. Thus, in addition to their usefulness for CMB studies, there is the potential to produce well-sampled spectral energy distributions. An important motivation for our work is to contribute to this undertaking with our high radio frequency flux density measurements.

Sources contained in CRATES are those most likely to contaminate {\it Planck} measurements of the CMB. Observations of these sources at 30~GHz using the OCRA-p receiver on the Toru\'n 32-m telescope, as described in this paper, could potentially be used to subtract the sources from the 30~GHz Low Frequency Instrument (LFI) maps, and will be useful for cross-comparison with the source lists generated by the {\it Planck} High Frequency Instrument (HFI). In the northern hemisphere, {\it Planck} will make its most sensitive observations at the north ecliptic pole, hence this is the best field in which to observe a sub-sample of the CRATES sources.

\section{Subsample selection}
The CRATES source sample consists of extragalactic sources ($|b| > 10 \ensuremath{^\circ}$) that have spectral indices between 1.4 and 4.8~GHz of $\alpha > -0.5$ (where $S \propto \nu^\alpha$), and have 4.8~GHz fluxes greater than 65 mJy. In the northern hemisphere the CRATES sample was selected from the 4.85~GHz GB6 catalogue \citep{1996Gregory} making use of 1.4~GHz flux densities from the NRAO VLA Sky Survey \citep[NVSS;][]{1998Condon} to select only the flat spectrum sources, i.e. those with spectral index $\alpha_{1.4}^{4.8} >-0.5$. CRATES is essentially a subset of the Cosmic Lens All-Sky Survey \citep[CLASS;][]{2003Myersa,2003Browne} combined with other surveys from the southern hemisphere. It should also be noted that at declinations above $75 \ensuremath{^\circ}$, where GB6 is incomplete, it may be missing significant numbers of sources; efforts to improve its completeness are ongoing \citep{2009Healey}. The complete sample contains $\sim$11~000 sources.

The subsample observed with OCRA has a series of additional selection criteria. We require that the sources:
\begin{itemize}
\item are closer than 35\ensuremath{^\circ} from the North Ecliptic Pole\\(located at RA $18^\mathrm{h}$ Dec $\sim$66.5\ensuremath{^\circ});
\item are at Dec $<75$\ensuremath{^\circ} (the GB6 survey limit);
\item are no closer than 15\ensuremath{^\circ} to the Galactic plane ($|b| > 15 \ensuremath{^\circ}$);
\item have a measured 8.4~GHz flux density.
\end{itemize}

This selection yields \changed{754} ``sources''. However, the CRATES catalog sometimes lists several components belonging to a single source, which would all lie within the 1.2~arcmin OCRA beam. For all CRATES entries with multiple components closer together than 1~arcmin we added their 8.4~GHz flux densities together. \changed{Most of the 149 extra components are very close to the main component (63 are within 0.1~arcmin and 120 are within 0.5~arcmin). They are also substantially lower in flux density than the main components (typically a few mJy at 8.4~GHz), and as such merging the multiple components together should negligibly affect the source spectral indices.} The position of the brightest component is used as the pointing position for our observations. Our final target-list contains 605 sources.

The RA range of the source sample is $9^\mathrm{h}~54^\mathrm{m}$ to $22^\mathrm{h}~6^\mathrm{m}$; the declination range is from $31^\circ~44^\mathrm{\prime}$ to $74^\circ~57^\mathrm{\prime}$. The 605 CRATES sources within this sample are shown in black in Figure \ref{fig:crates_sources_aitoff}, with all of the rest of CRATES sources shown in grey.

\section{OCRA-p observations}
OCRA-p is a two-beam 30~GHz instrument mounted on the 32-m Toru\'n telescope. It is the first instrument of the OCRA programme \citep{2000Browne}. This two-beam pseudo-correlation receiver is based upon the Planck LFI receiver chain, and is also similar in concept to the WMAP 23~GHz receiver. With an overall system temperature of 40~K, and a bandwidth of 6~GHz, the receiver has a sensitivity of 6~mJy~s$^{1/2}$. It has previously been used to observe strong radio sources \citep{2007Lowe}, planetary nebulae \citep{2009Pazderska}, weak radio sources \citep{2009Gawronski} and clusters of galaxies by the Sunyaev-Zel'dovich Effect \citep{2006Lancaster,2010Lancaster}. The OCRA-p data reduction process is described in detail in \citet{2009Gawronski} and \citet{2009Peela}.

In order to optimise the observing efficiency an expected flux density of each source at 30~GHz was calculated from an extrapolation from the 4.8 and 8.4~GHz flux densities. Those sources that were predicted to be stronger than 100~mJy were observed using cross-scan measurements, first in elevation then in azimuth, as described in \citet{2007Lowe} and \citet{2009Gawronski}. As cross-scan measurements provide simultaneous pointing corrections for the telescope, these are robust flux density measurements. A different strategy is used for sources $<100$~mJy, which may be too faint for reliable cross-scan observations in the face of atmospheric fluctuations. For these objects, following a cross-scan pointing observation on a nearby bright and unresolved source closer than 4$^\circ$ to the fainter source, a set of ``on-off'' measurements are carried out, where the two beams of the instrument are alternately pointed towards the source position, as described in \citet{2009Gawronski}. Any source not detected using the cross-scan method was re-observed using on-off measurements. This combination of the two observational techniques provides the most efficient route to flux density measurements with OCRA-p.

As OCRA-p is sensitive to a single linear polarisation, measurements of the flux density of polarised sources will vary depending on the hour angle of the observation. However, as flat spectrum radio sources are only weakly polarised at high frequencies ($\sim$ 3 per cent; see e.g. \citealp{2009Jackson}) this will not be a significant effect compared to the intrinsic source variability and the measurement uncertainty.

We use the planetary nebula NGC~7027 for primary flux density calibration. This has been measured at 33~GHz by \citet{2008Hafez}, who report a flux density of $5.39 \pm 0.04$~Jy at an epoch of 2003.0 with a secular decrease of $-0.17 \pm 0.03$~per~cent per year. Extrapolating this to \changed{2009.0} yields a flux density of $5.34 \pm 0.04$~Jy, which can then be scaled to 30~GHz using the quoted spectral index of -0.119 to give $\changed{5.40} \pm 0.04$~Jy. We note that observations of NGC~7027 by \citet{2008Zijlstra} are consistent with \citet{2008Hafez}. \changed{The absolute calibration used by \citet{2008Hafez} was the 5-year WMAP brightness temperature for Jupiter of 146.6$\pm$0.8~K \citep{2009Hill}. We note that there is a debate about the effect of potential timing errors, which could affect the flux density scale at the $\sim$0.2\% level \citep[][and references therein]{2010Boud}.}

Secondary flux density calibration is carried out using a signal from a noise diode measured after each source observation. We correct for the changes in system gain as a function of elevation using a series of measurements of NGC~7027 at different elevations. We also correct for atmospheric absorption based on measurements of the system temperature at zenith and 60$\ensuremath{^\circ}$ from the zenith.

Observations commenced in November 2008 and were completed in June 2010. A total of 4,105 separate observations have been made -- an average of 6.8 measurements per source, with each source having at least 3 good measurements. 

Cross-scan measurements where the amplitudes recorded from the two beams disagree by greater than 20 per cent, or where the widths of the peaks disagree by greater than 40 per cent, are flagged and removed from the analysis. Measurements that are obviously affected by poor weather are manually flagged. We also automatically flag any on-off measurements with an error on the measurement that is both over 7~mJy and greater than 15 per cent of the flux density of the measurement. Additionally, measurements that are clearly erroneously low are manually flagged as these are likely to have been caused by the telescope not being positioned accurately on the source. In total, 1,104 measurements have been flagged; 3001 measurements, or an average of 5 per source, are used in the final analysis.

\changed{The reported errors on the OCRA-p flux density measurements were calculated using the weighted standard deviation of the unflagged measurements of each source, where the weights were determined using the uncertainty in the least-squares fits to the cross-scan measurements or the scatter of the 1~second samples within the on-off measurements (see \citealp{2009Gawronski,2009Peela} for more details).}

To compensate for the effects of small random pointing errors in the on-off measurements, we have increased the measured flux density for on-off measurements for the CRATES sample by 5 per cent. This is an empirical number determined from a sample of sources where we made measurements using both methods. We also quadratically add 5 per cent of the measured flux density to the measurement error to account for the pointing uncertainty during on-off measurements.

As with the VSA sources \citep{2009Gawronski}, the final error on the flux density for each source is calculated by $\sigma = \sqrt{\sigma_\mathrm{meas}^2 + (0.08 S)^2}$ where the 8 per cent of the flux density takes into account the uncertainty due to calibration, atmospheric and gain-elevation corrections and other atmospheric effects (e.g. $1/f$-like fluctuations).

\section{Flux densities} \label{sec:crates_source_fluxes}
The flux densities of the CRATES source subsample as measured with OCRA-p are listed in Table \ref{tab:crates_fluxes}. The 1.4, 4.8 and 8.4~GHz flux densities from the CRATES source catalogue are also listed. We have made a machine readable copy of the flux density table, as well as plots of the measurements of each source over time and the aggregated source spectra between 26~MHz and 150~GHz, available online at \url{http://www.jodrellbank.manchester.ac.uk/research/ocra/crates/}.

From an inspection of NVSS \citep{1998Condon}, FIRST \citep{1995Becker} and CLASS \citep{2003Myersa,2003Browne} data it is clear that the majority of the CRATES sources are unresolved. A number do, however, show extension and/or multiple components; these are marked with an ``e'' in Table \ref{tab:crates_fluxes}. Sources where multiple components listed in CRATES have been merged together are denoted ``NC'' where ``N'' is the number of components. WMAP sources \citep{2010Gold} are marked with a ``w'', CJF sources \citep{1996Taylor,2007Lowe} are marked with a ``c'' and Fermi point sources \citep{2010Abdoa} are marked with an ``f'' (see the next section). Potential Gigahertz Peaked Spectrum (GPS) sources are marked with a ``g'' (see Section 6). We note that there are two CLASS lens systems within the sample -- J1609+6532 and J1938+6648 \citep{2003Browne}.

There are 42 sources that appear variable within the OCRA measurements; these are marked with a ``v'' in Table \ref{tab:crates_fluxes}. Sources were identified as variable if they are reasonably strong ($>$20~mJy) and have two measurements which are 25 per cent higher than the mean. Although sources with two measurements that are 25 per cent lower than the mean could also be variable sources, these will be more contaminated by any sources with measurements that suffer from bad telescope pointing (and hence a reduced measured flux density) that have not been flagged. As such, we do not include these in the list of variable sources.

\section{Comparison with other measurements}
\subsection{CJF Sources}
\begin{figure}
\centering
\includegraphics[scale=0.33]{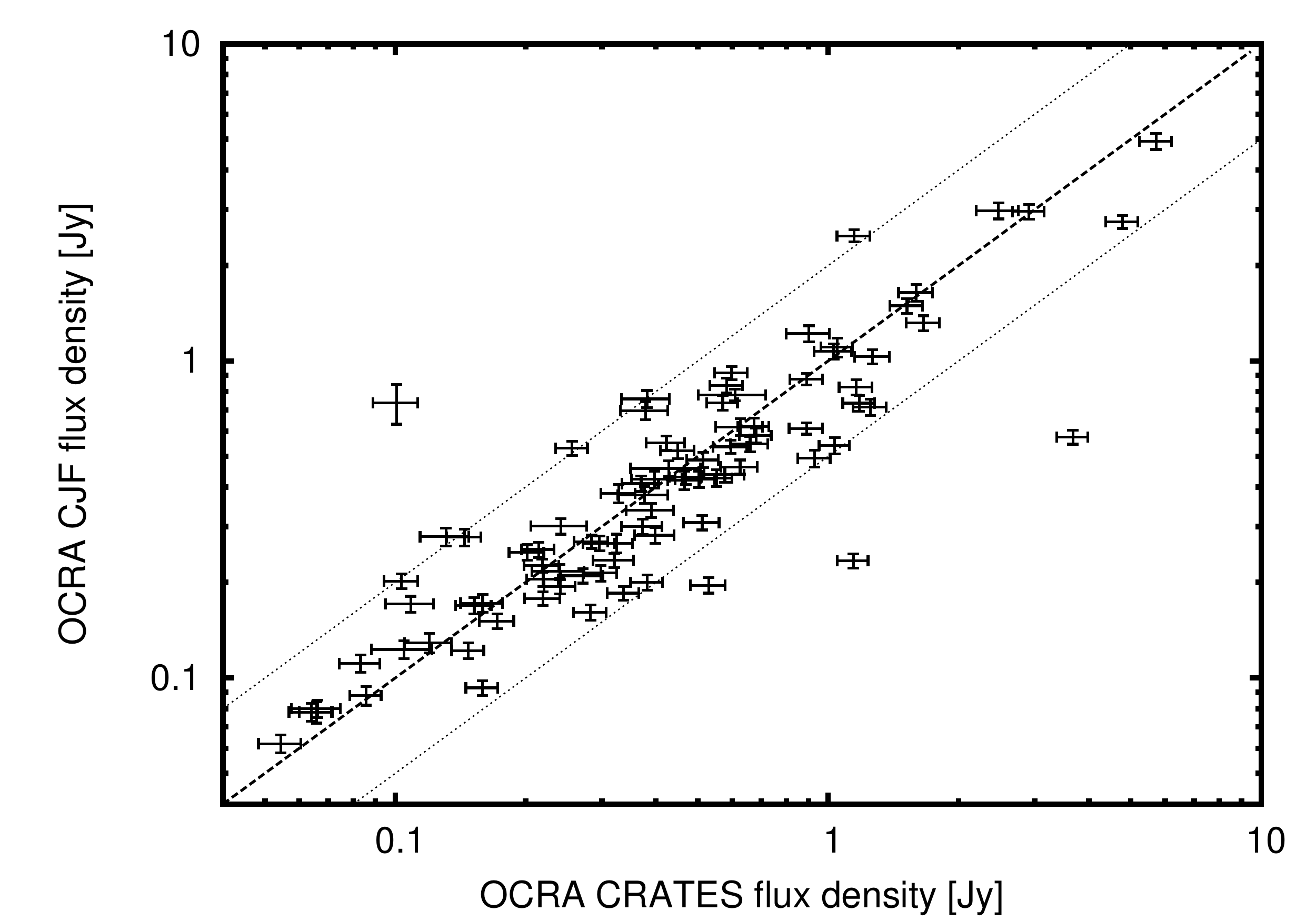}
\caption{Comparison of common sources between CJF and CRATES. The dashed line is $x=y$; the dotted lines indicate variability of a factor of two. The measurements broadly agree, with scatter ascribed to variability. Four sources particularly stand out as having varied significantly in flux density (see text).}
\label{fig:comparison_cj_crates}
\end{figure}

There are 87 sources in common between the CRATES subsample and the OCRA-p measurements of CJF sources reported in \citet{2007Lowe}; the flux densities of these are plotted in Figure \ref{fig:comparison_cj_crates}. There is broad agreement between the two, although there is a large degree of scatter.

Four of the sources have particularly discrepant flux densities (by a factor greater than 2), which can be certainly ascribed to intrinsic variability. J1849+6705 has increased in flux density from 575$\pm$29~mJy in \citet{2007Lowe} to 3675$\pm$301~mJy. J1852+4855 was 196$\pm$11~mJy, and is now 530$\pm$49~mJy. J2006+6424 was 234$\pm$12~mJy; it is now 1143$\pm$93~mJy. Finally, J0954+7435 was 738$\pm$106~mJy and has now decreased to 101$\pm$12~mJy. This last type of source behaviour could present significant problems for non-contemporaneous measurements of point sources for subtraction from CMB maps, as the sources that have faded will be difficult to recover from later surveys unless they brighten again in the future.

Additionally, J1642+6856 has also increased significantly, from 2.75$\pm$0.13~Jy to 4.79$\pm$0.41~Jy, with a large amount of scatter within our OCRA measurements over the last year implying ongoing variability.

\subsection{WMAP sources}
\begin{table}
\begin{tabular}{c|c|c|c|l}
{\bf Name} & {\bf OCRA} & {\bf WMAP K} & {\bf WMAP Ka} & {\bf Notes}\\
& {\bf 30~GHz} & {\bf 23~GHz} & {\bf 33~GHz} & \\
& \multicolumn{3}{|c|}{{\bf Flux density/Jy}} & \\
\hline
J1048+7143 & 1.15$\pm$0.10 & 1.21$\pm$0.61 & 1.11$\pm$0.73 & \\ 
J1101+7225 & 0.93$\pm$0.08 & 0.92$\pm$0.16 & 0.91$\pm$0.18 & a\\ 
J1302+5748 & 0.38$\pm$0.05 & 0.81$\pm$0.17 & 0.67$\pm$0.18 & a\\ 
J1343+6602 & 0.32$\pm$0.03 & 0.70$\pm$0.17 & 0.34$\pm$0.13 & Pa\\ 
J1344+6606 & 0.37$\pm$0.04 & as above & as above & \\
J1419+5423 & 0.61$\pm$0.11 & 0.86$\pm$0.18 & 0.89$\pm$0.25 & a\\ 
J1436+6336 & 1.04$\pm$0.08 & 0.68$\pm$0.23 & 0.50$\pm$0.28 & a\\ 
J1549+5038 & 0.57$\pm$0.05 & 0.85$\pm$0.14 & 0.64$\pm$0.11 & \\ 
J1604+5714 & 0.49$\pm$0.09 & 0.81$\pm$0.12 & 0.77$\pm$0.17 & a\\ 
J1625+4134 & 0.45$\pm$0.04 & 0.96$\pm$0.06 & 0.78$\pm$0.14 & \\ 
J1635+3808 & 2.47$\pm$0.28 & 3.72$\pm$0.86 & 4.06$\pm$1.08 & \\ 
J1637+4717 & 1.27$\pm$0.12 & 1.10$\pm$0.26 & 1.12$\pm$0.26 & \\ 
J1638+5720 & 1.67$\pm$0.15 & 1.67$\pm$0.43 & 1.68$\pm$0.48 &\\ 
J1642+3948 & 5.72$\pm$0.48 & 6.60$\pm$0.90 & 5.95$\pm$0.94 & \\ 
J1642+6856 & 4.79$\pm$0.41 & 1.82$\pm$0.68 & 1.85$\pm$0.85 & \\ 
J1648+4104 & 0.45$\pm$0.05 & 0.67$\pm$0.08 & 0.82$\pm$0.20 & a\\ 
J1653+3945 & 0.90$\pm$0.10 & 1.25$\pm$0.12 & 1.16$\pm$0.16 & a\\ 
J1657+5705 & 0.51$\pm$0.04 & 0.41$\pm$0.10 & -- &\\ 
J1658+4737 & 0.68$\pm$0.06 & 1.16$\pm$0.10 & 1.16$\pm$0.10 & \\ 
J1700+6830 & 0.39$\pm$0.05 & 0.32$\pm$0.07 & 0.51$\pm$0.08 & \\ 
J1701+3954 & 0.52$\pm$0.04 & 0.58$\pm$0.09 & 0.77$\pm$0.20 & \\ 
J1716+6836 & 0.60$\pm$0.05 & 0.58$\pm$0.11 & 0.51$\pm$0.16 & \\ 
J1727+4530 & 1.25$\pm$0.11 & 0.89$\pm$0.44 & 0.94$\pm$0.52 & \\ 
J1734+3857 & 1.05$\pm$0.09 & 1.20$\pm$0.11 & 1.26$\pm$0.24 & \\ 
J1735+3616 & 0.51$\pm$0.04 & 0.71$\pm$0.09 & 0.70$\pm$0.16 & \\ 
J1739+4737 & 0.67$\pm$0.05 & 0.86$\pm$0.12 & 0.77$\pm$0.20 & \\ 
J1740+5211 & 1.16$\pm$0.10 & 1.22$\pm$0.26 & 1.18$\pm$0.20 & \\ 
J1748+7005 & 0.47$\pm$0.04 & 0.50$\pm$0.15 & 0.63$\pm$0.16 & \\ 
J1753+4409 & 0.42$\pm$0.04 & 0.63$\pm$0.21 & 0.61$\pm$0.13 & a\\ 
J1800+3848 & 0.89$\pm$0.08 & 0.91$\pm$0.11 & 0.72$\pm$0.27 & a\\ 
J1801+4404 & 1.18$\pm$0.10 & 1.36$\pm$0.26 & 1.52$\pm$0.27 & \\ 
J1806+6949 & 1.52$\pm$0.13 & 1.42$\pm$0.14 & 1.36$\pm$0.22 & \\ 
J1812+5603 & 0.27$\pm$0.03 & 0.21$\pm$0.11 & 0.30$\pm$0.22 & \\ 
J1824+5651 & 1.60$\pm$0.14 & 1.47$\pm$0.26 & 1.24$\pm$0.18 & \\ 
J1835+3241 & 0.41$\pm$0.04 & 0.81$\pm$0.09 & 0.81$\pm$0.15 &\\ 
J1842+6809 & 0.95$\pm$0.09 & 1.29$\pm$0.27 & 1.40$\pm$0.30 & a\\ 
J1849+6705 & 3.67$\pm$0.30 & 1.63$\pm$0.66 & 1.81$\pm$0.78 & a\\ 
J1927+6117 & 0.58$\pm$0.05 & 0.95$\pm$0.26 & 0.94$\pm$0.22 & \\ 
J1927+7358 & 2.91$\pm$0.25 & 3.57$\pm$0.27 & 3.33$\pm$0.36 & \\ 
J2007+6607 & 0.63$\pm$0.06 & 0.78$\pm$0.09 & 0.57$\pm$0.08 & \\ 
J2009+7229 & 0.60$\pm$0.05 & 0.64$\pm$0.12 & 0.78$\pm$0.16 & \\ 
J2015+6554 & 0.36$\pm$0.03 & 0.75$\pm$0.14 & 0.79$\pm$0.15 & a\\ 
\end{tabular}
\caption{Comparison of 42 CRATES sources measured with OCRA-p that are also in the WMAP 7-year point source catalogue \citep{2010Gold}. The flux densities are given in Jy. Two sources -- J1343+6602 and J1344+6606 -- are combined in WMAP (denoted with ``P'').}
\label{tab:crates_wmap}
\end{table}

\begin{figure}
\centering
\includegraphics[scale=0.33]{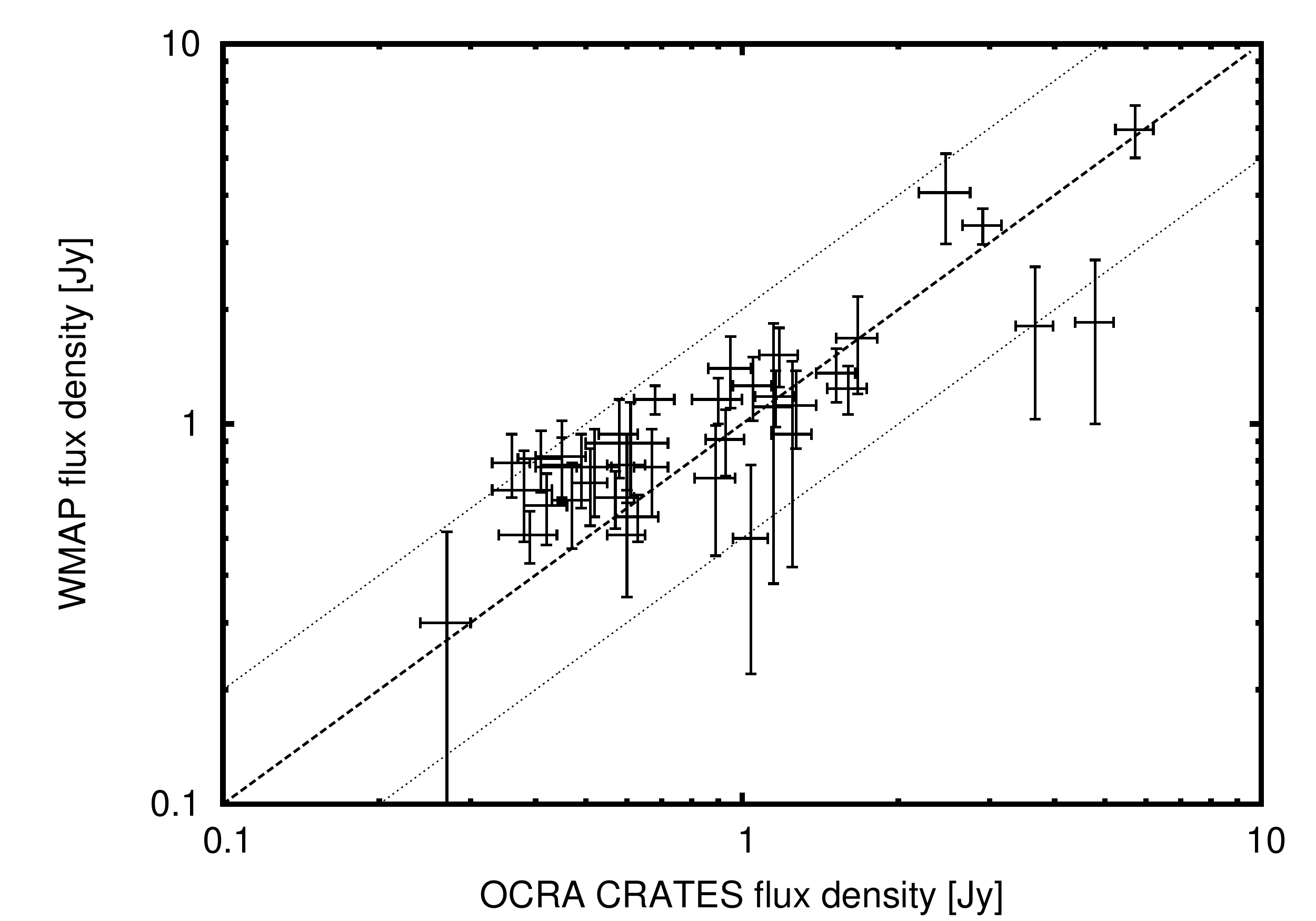}\\
\caption{Comparison of OCRA-p and WMAP Ka-band measurements of common sources (as per Table \ref{tab:crates_wmap}). The paired sources, and the source with the missing flux density, have not been included. The dashed line is $x=y$; the dotted lines indicate variability of a factor of two.}
\label{fig:crates_comparison_wmap}
\end{figure}
The subsample also has 42 sources in common with the 7-year WMAP point source catalogue \citep{2010Gold}. These are listed in Table \ref{tab:crates_wmap}, which gives the OCRA measurement and the WMAP K (23~GHz) and Ka-band (33~GHz) flux densities, and \changed{a comparison of the OCRA and WMAP Ka-band flux densities is plotted in} Figure \ref{fig:crates_comparison_wmap}. Where there are multiple identifications for the WMAP sources (as given in \citealp{2010Gold}) the sources are marked with an ``a''. Two CRATES sources -- J1343+6602 and J1344+6606 -- are \changed{separated} by only 3.8~arcmin. \citet{2010Gold} list a single combined flux density for these sources. We also note that J1657+5705 had a Ka-band flux density of 0.6$\pm$0.09 Jy in \citet{2009Wright}, but the flux density at this frequency is omitted in \citet{2010Gold}.

The uncertainties of the WMAP flux densities were calculated from the scatter between the yearly WMAP measurements, and as such they combine the Gaussian and systematic errors as well as the source variability. The uncertainties on the OCRA-p measurements were calculated by the same method, although the measurements were taken over a shorter time period. There are 5 sources with a large uncertainty in both the WMAP and OCRA measurements -- J1635+3808, J1642+3948, J1642+6867, J1849+6705, J1927+7358 -- implying that these are variable on short ($\sim$year-long) timescales. Three sources have a large uncertainty in WMAP but not in the OCRA measurements -- J1048+7143, J1638+5720 and J1727+4530 -- implying that these are variable on longer timescales only, or were otherwise stable during the OCRA-p measurements.

At the lowest flux densities, we note that the WMAP flux densities are systematically higher than those from OCRA-p. These sources have likely been Eddington biased into the WMAP sample either by noise bias, variability or due to extra flux density seen by the large WMAP beam from nearby point sources, diffuse foreground emission or the CMB -- a 50~$\upmu$K fluctuation in the CMB corresponds to $\sim$0.25~Jy at Ka-band in WMAP \citep{2010Jarosik}.


\begin{figure}
\centering
\includegraphics[scale=0.33]{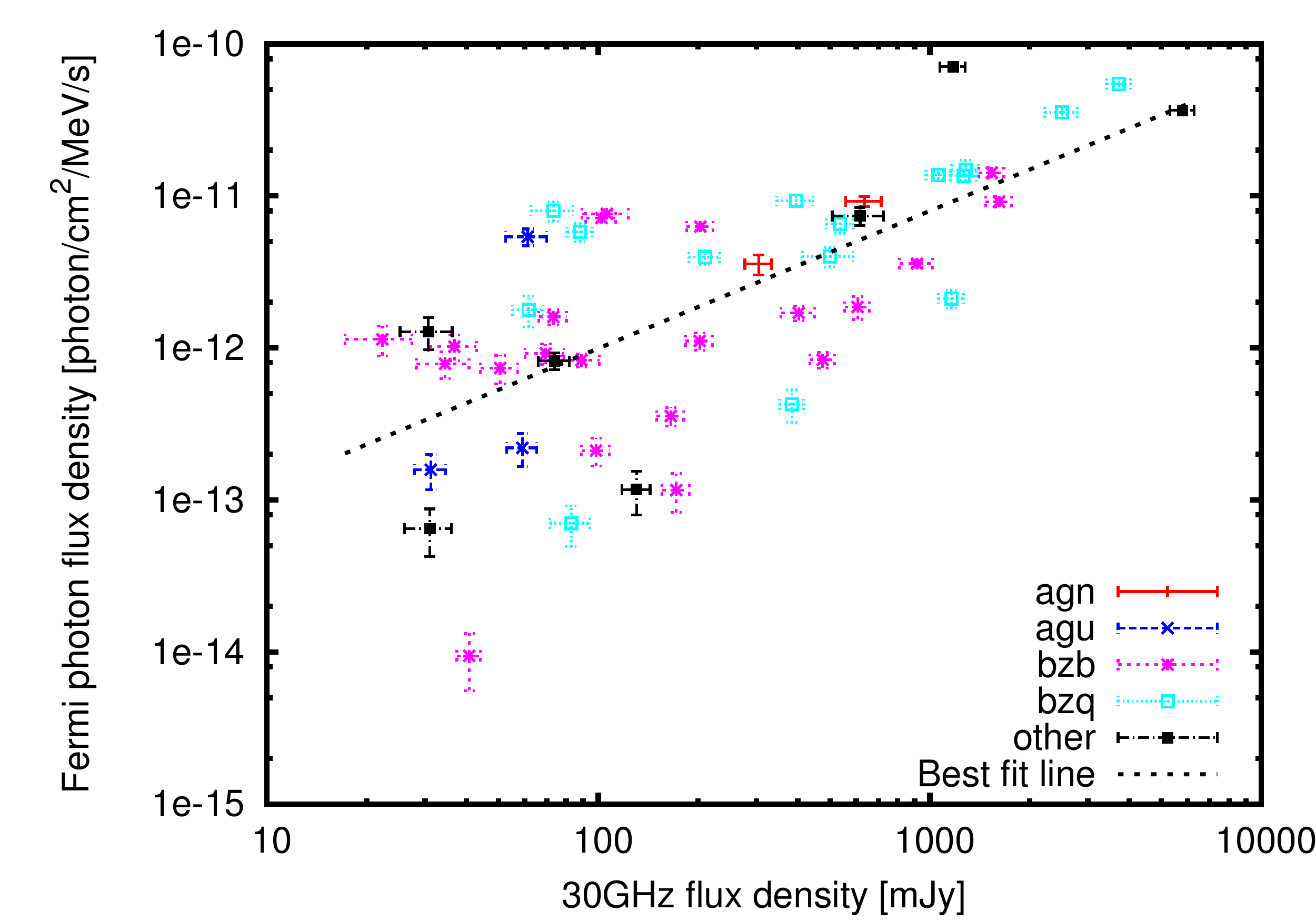}\\
\caption{Comparison of the 30~GHz flux density from OCRA with the Fermi flux density for common sources. There is a surprisingly good correlation, shown by the line of best fit.}
\label{fig:crates_comparison_fermi}
\end{figure}

\subsection{Fermi sources}
We also carry out a cross-correlation of the CRATES sources with the Fermi-LAT 1 year point source catalogue \citep[1FGL; ][]{2010Abdoa} and find 48 sources in common. The Fermi flux densities compared with the 30~GHz OCRA-p measurements are shown in Figure \ref{fig:crates_comparison_fermi}. We find a good correlation between the radio and gamma ray flux densities, most clearly indicated by the lack of sources with high radio flux densities and low Fermi flux densities. A Spearman-rank correlation test gives $\rho = 0.71$ indicating a strong correlation (significant at greater than the 99.5\% level); applying a best fit line of form $\log_{10}(S_{30}) = m \log_{10}(S_\mathrm{F}) + c$ gives $m=0.90\pm0.02$, $c=-13.8\pm0.1$. We split the sample up into different population classes using the classifications given in the 1FGL catalogue (i.e. BL Lac, quasar, etc.), and find no obvious dependency on population.

\begin{figure}
\centering
\includegraphics[scale=0.3]{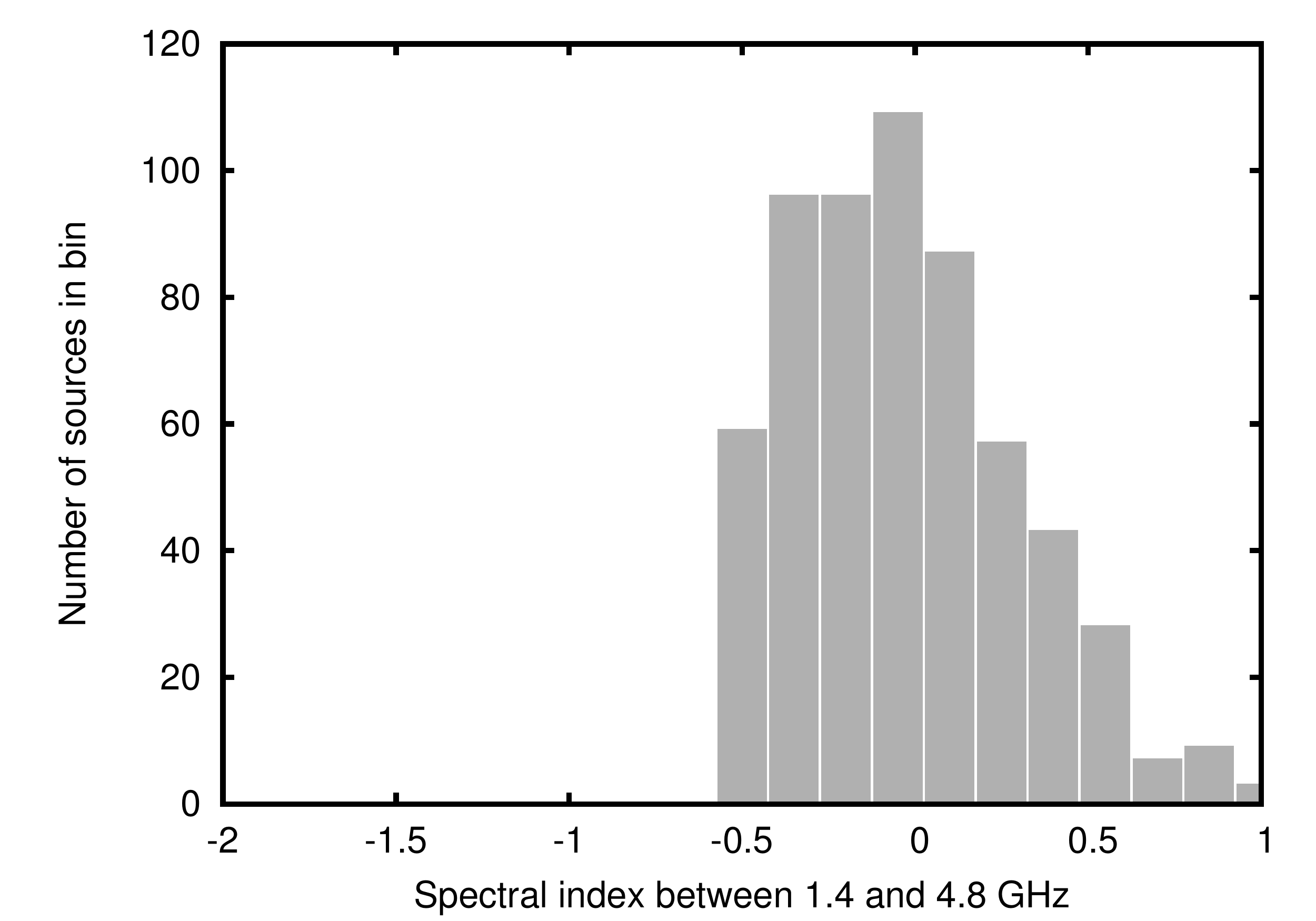}
\includegraphics[scale=0.3]{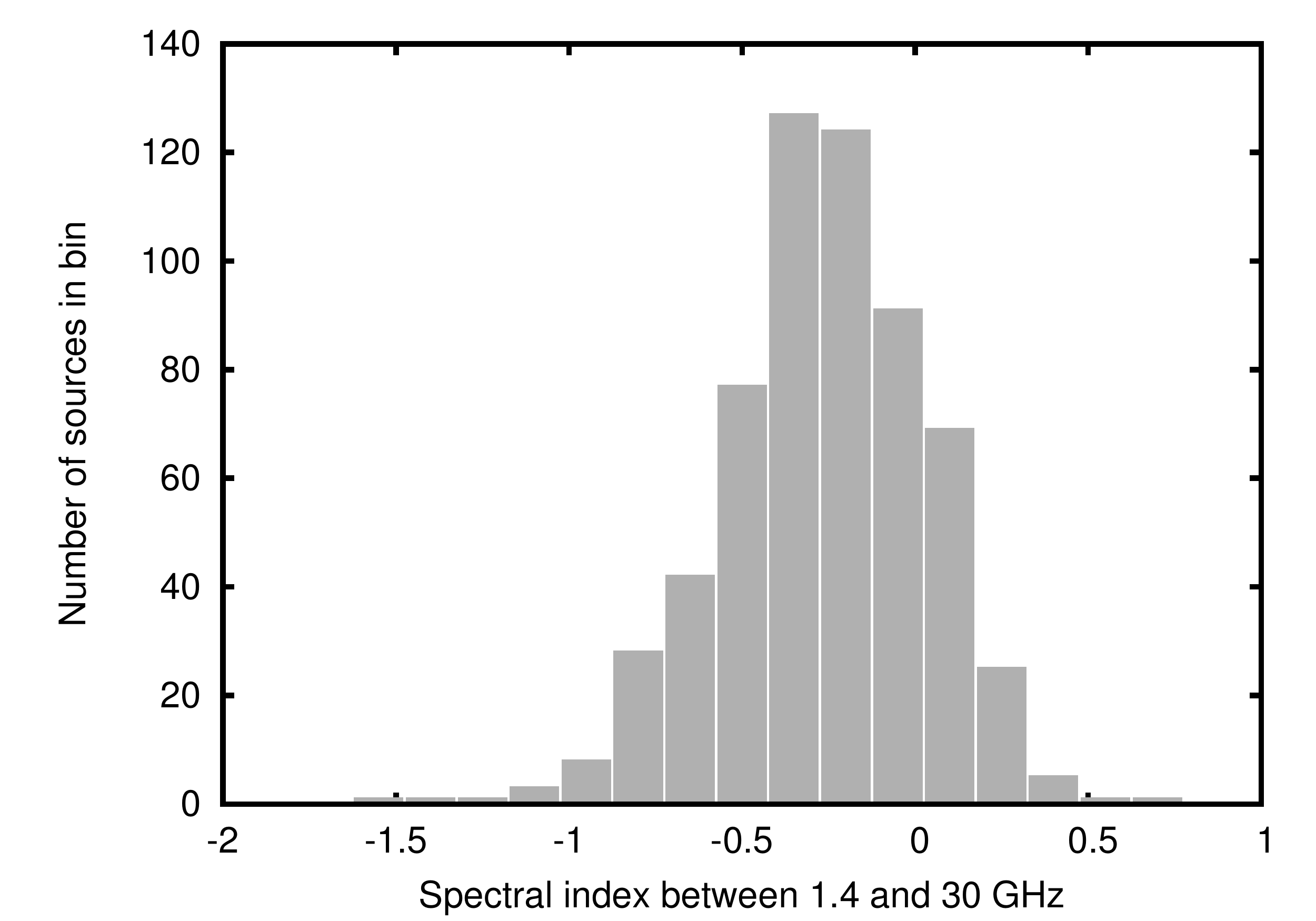}\\
\includegraphics[scale=0.3]{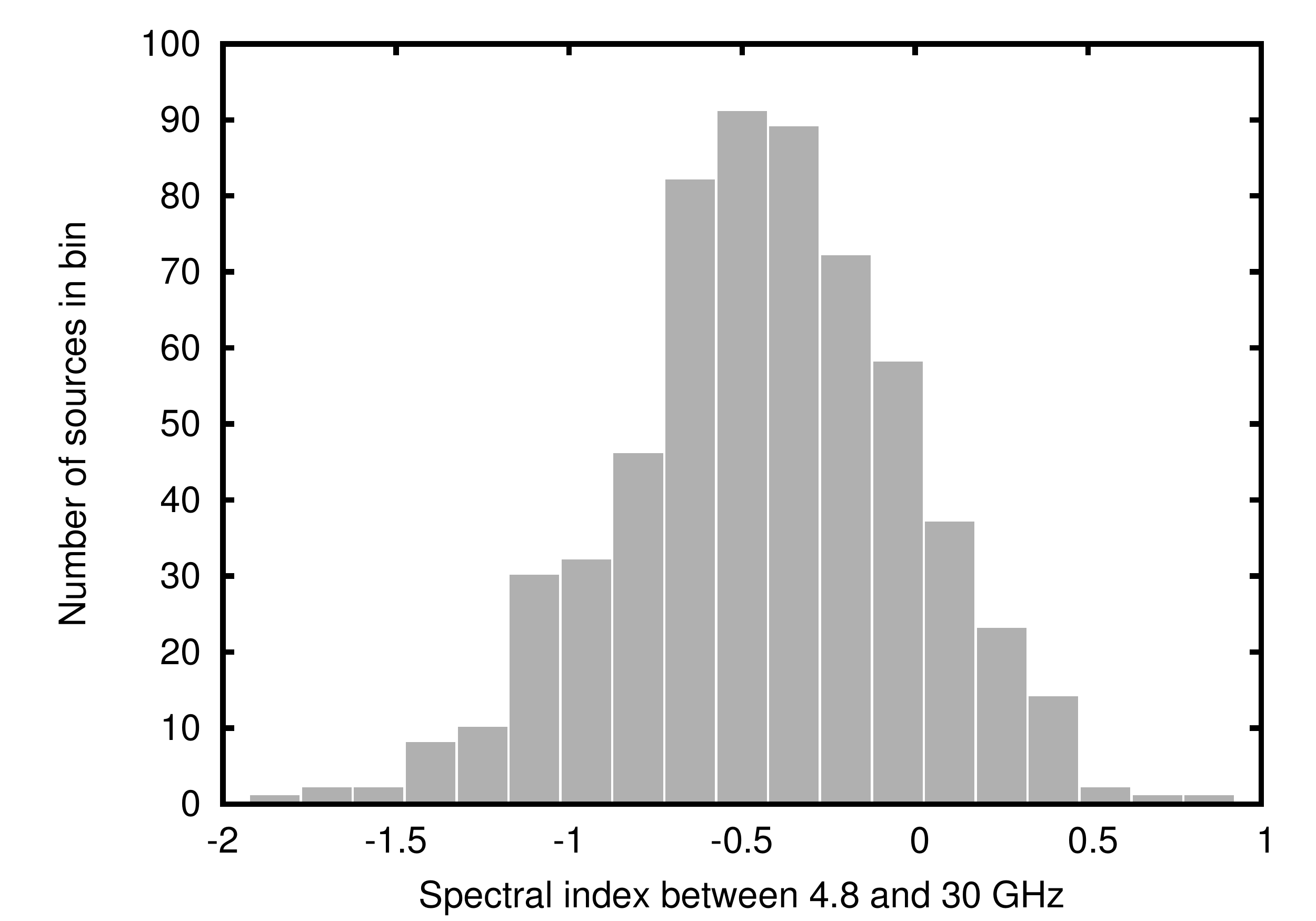}
\includegraphics[scale=0.3]{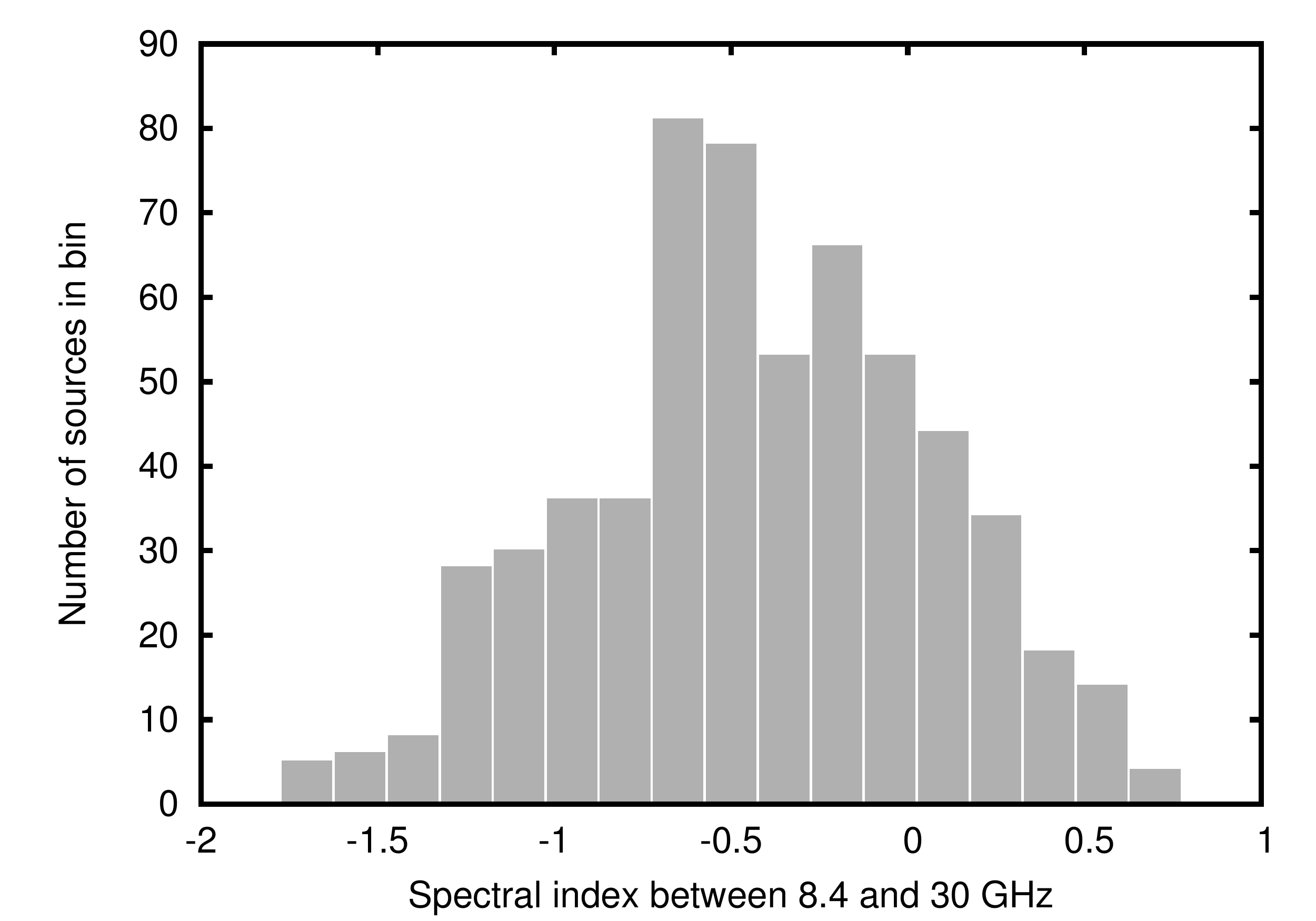}\\
\caption[Spectral index distributions for the CRATES source sample] {The spectral index distributions for the sources in the CRATES sample observed with OCRA-p. The hard cut-off in the first panel is due to the selection criteria of $\alpha_{1.4}^{4.8}>-0.5$.}
\label{fig:crates_spectralindex_histogram}
\end{figure}

\begin{figure}
\centering
\includegraphics[scale=0.33]{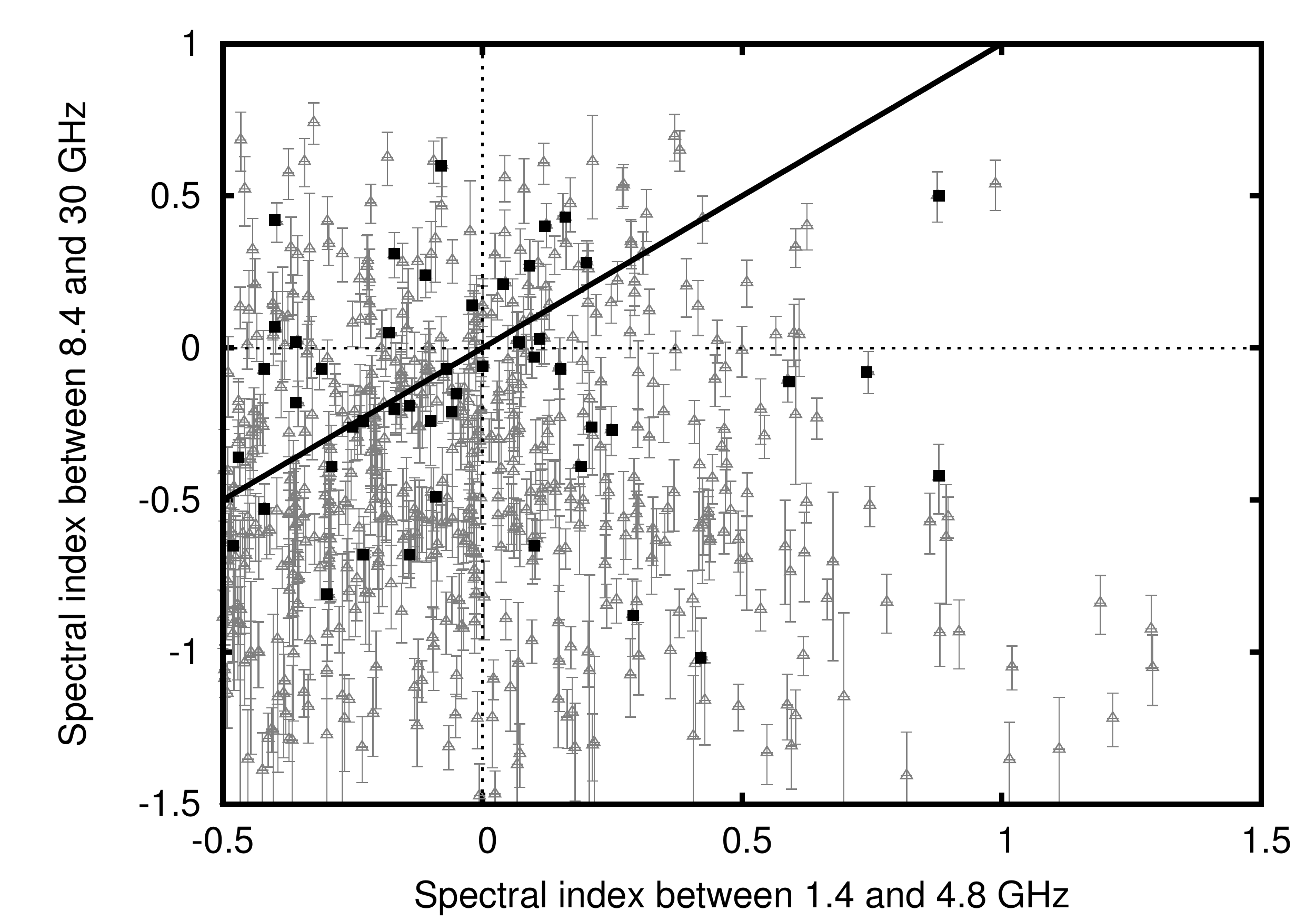}\\
\caption{A 2-colour diagram for the CRATES source sample showing the spectral index from 1.4 to 4.8~GHz compared with the spectral index from 8.4 to 30~GHz. The solid line indicates $x=y$; 75 per cent of sources have steepened and lie below this line. The darker points are sources also detected in the 1FGL \citep{2010Abdoa}.}
\label{fig:crates_histogram_2colour}
\end{figure}

The above result is similar to that from the comparison between Fermi sources and the AT20G survey by \citet{2010Mahony} and \citet{2010Ghirlandaa, 2010Ghirlandab}, and also the comparison with the CRATES 8.4GHz measurements by \citet{2010Giroletti}. We find a significantly higher correlation and slope than those comparisons, however we caution that sample bias will be present as CRATES is not a blind survey. It is also important to take the variability of both the gamma-ray and radio flux densities into account when interpreting this result, although this issue is lessened in this sample compared with AT20G due to the reduced separation in time between the gamma ray and radio measurements. In particular, \citet{2010Ghirlandab} have shown that the slope of the relationship is strongly affected by selection effects and variability. Because we have not included sources with upper limits on their gamma-ray flux densities, the true slope of the correlation is uncertain but is likely to be steeper.

\section{Spectral index distributions}

The spectral index distributions for the CRATES sources between 1.4 and 4.8; 1.4 and 30; 4.8 and 30 and 8.4 and 30~GHz are shown in Figure \ref{fig:crates_spectralindex_histogram}. The effects of the selection criterion of $\alpha_{1.4}^{4.8}>-0.5$ is obvious in the first of these. However, when flux densities measured at other frequencies are considered the effect of the selection cut becomes blurred. The mean and \changed{standard deviation} for $\alpha_{1.4}^{4.8}$ is $0.00\pm0.38$; this changes to $-\changed{0.27}\pm$0.29 for $\alpha_{1.4}^{30}$, $-\changed{0.45}\pm0.44$ for $\alpha_{4.8}^{30}$ and $-\changed{0.46}\pm0.56$ for $\alpha_{8.4}^{30}$. This reflects the general steepening of the source spectra at high frequencies, as seen by e.g. \citet{1966Kellerman, 2006Ricci, 2007Lowe, 2007Massardi}. \changed{We caution} that the sample selection excludes sources with $\alpha_{1.4}^{4.8}<-0.5$ and that, since the primary selection frequency is 5~GHz, there is a bias against those sources that have spectra which flatten or \changed{turn up} at frequencies above this.

The 2-colour diagram depicting $\alpha_{1.4}^{4.8}$ vs. $\alpha_{8.4}^{30}$ is shown in Figure \ref{fig:crates_histogram_2colour}. From this, we find that:
\begin{itemize}
\item 75 per cent of the sources have steepened at $\alpha_{8.4}^{30}$ compared with $\alpha_{1.4}^{4.8}$;
\item 34 sources (\changed{6} per cent) appear to be Gigahertz-peaked sources that peak between 4.8 and 8.4~GHz (defined by $\alpha_{1.4}^{4.8} > 0.5$ and $\alpha_{8.4}^{30} < -0.5$; however see below);
\item \changed{63} sources (\changed{10} per cent) are flat or rising ($\alpha_{1.4}^{4.8} > 0$ and $\alpha_{8.4}^{30} > 0$;
\item \changed{62} sources (\changed{10} per cent) are inverted ($\alpha_{1.4}^{4.8} < 0$ and $\alpha_{8.4}^{30} > 0$).
\end{itemize}

\begin{table}
\begin{tabular}{c|c|l}
{\bf Name} & {\bf Cross ID} & {\bf Notes}\\
\hline
J0954+7435 & VCS J0954+7435 &d?\\
J1143+6619 & --&\\
J1143+6633 & VCS J1143+6633 & cj?\\
J1144+6844 & -- &\\
J1210+6422 & VIPS J12105+6422 & d\\
J1245+7117 & -- &\\
J1247+7124 & VCS J1247+7124 & s?; C(M99)\\
J1305+5836 & -- & \\
J1333+6737 & -- &\\
J1336+7437 & -- &\\
J1338+6632 & -- &\\
J1436+4820 & -- &\\
J1443+6332 & VIPS J14439+6332 & d\\
 & VCS J1443+6332 & d?\\
J1446+6043 & -- &\\
J1457+6357 & VIPS J14577+6357 & cj?\\
J1525+6751 & -- & GPS(S98)\\
J1545+3941 & VCS J1545+3941&GPS(M99)\\ 
J1549+6134 & -- &\\
J1558+6521 & -- &\\
J1613+4223 & VIPS J16130+4223 & s; C(M99)\\
J1613+6329 & -- &\\
J1614+5826 & -- &\\
J1619+7226 & -- &\\
J1624+6259 & -- &\\
J1628+4734 & VIPS J16286+4734 & cj; GPS(M99)\\
& VCS J1628+4734 & \\
J1630+6659 & -- &\\
J1720+3604 & -- &\\
J1741+4751 & VIPS J17415+4751 & cj \\ 
& VCS J1741+4751 & \\
J1756+5748 & VIPS J17560+5748 & t; GPS(M99)\\
& VCS J1756+5748 (t) & \\
J1757+4757 & VIPS J17574+4757 & s\\
J1800+3848 & VCS J1800+3848 & s; GPS(L07)\\
J1800+7325 & -- &\\
J1803+6433 & -- &\\
J1909+6241 & -- &\\
J1916+6053 & -- &\\
J1919+7240 & -- &\\
J1945+7055 & VCS J1945+7055 & c; GPS(S98)\\
J1955+6500 & -- &\\
\end{tabular}
\caption{Candidate GPS sources, with VIPS \citep{2007Helmboldt} and VLBA Calibrator Survey (VCS; \citealp{2008Petrov}) identifications and notes where the sources have previously been considered as Gigahertz-Peaked (``C'' for Candidate, ``GPS'' for confirmed) by \citet[][S98]{1998Snellen}, \citet[][M99]{1999Marecki} and \citet[][L07]{2007Lowe}, and classifications of the morphology where interferometric maps are available (``s'' for single, ``d'' for double, ``t'' for triple, ``cj'' for core-jet and ``c'' for complex).}
\label{tab:crates_gps}
\end{table}

We caution that, due to the high resolution of the 8.4~GHz interferometric observations compared with the other measurements, sources that are extended or have multiple components will likely have underestimated flux densities at 8.4~GHz. This will have increased the number of sources with apparently rising spectra between 8.4 and 30~GHz. There are also some effects in the spectra due to the different resolutions and observational techniques at 1.4 and \changed{4.8~GHz: e.g., although} J1818+4916 and J1821+3602 have GPS-like spectra in the catalogue, the 1.4GHz flux densities do not include \changed{contributions} from all the extended emission of these steep spectrum \changed{sources. Put} another way, the GB6 observations overestimate the flux density of the compact components of these objects.

The sources cross-identified with the 1FGL catalogue \citep{2010Abdoa} are shown \changed{as black squares} in Figure \ref{fig:crates_histogram_2colour}. There is a tendency \changed{for these to have a flatter} spectrum between 8.4 and 30~GHz than the rest of the sources, qualitatively in agreement with \citet{2010Mahony} and \citet{2010Ghirlandaa}.

From a visual inspection of the spectra of the observed sources, combining the CRATES catalogue measurements with the OCRA-p measurements, and also available measurements in the literature (via the NASA/IPAC Extragalactic Database), we identify 38 candidate GPS sources, which are listed in Table \ref{tab:crates_gps}. We also identify a further 29 potential GPS sources, which are marked in Table \ref{tab:crates_fluxes} with ``G?''.

\citet{1998ODea} and \citet{2007Lowe} found that $\sim$ 10 per cent of their 5~GHz-selected sources were GPS sources. We find fewer candidate GPS sources than this (\changed{6} per cent), but a similar percentage of combined candidate and potential GPS sources (11 per cent). We ascribe this to the difficulty in determining whether a source is GPS or not with fewer data points in the source spectra than is the case for the \citet{1998ODea} and \citet{2007Lowe} samples. Additionally, variable sources may appear to have GPS spectra as the measurements at each frequency have been taken at different epochs; as such, simultaneous measurements at different frequencies will be important to confirm whether these candidates are indeed GPS sources. Finally, there are some sources that have a rising spectra at all frequencies; these could be GPS sources that peak at  frequencies greater than 30~GHz.

\section{Conclusions}
We have measured flux densities at 30~GHz for 605 flat-spectrum radio sources from the CRATES sample around the North Ecliptic Pole using OCRA-p and the 32-m \changed{Toru\'n} telescope. These observations follow from those of the CJF sources by \citet{2006Lowe} and \citet{2007Lowe}, extending the work to lower flux densities. Where sources are in common, we find reasonable agreement between the measurements presented here and those by \citet{2007Lowe} and also WMAP \citep{2010Gold}. However, a number of sources display variability between the three sets of observations, and also within the present measurements. Such variability implies that simple source subtraction strategies in CMB studies can be subject to considerable uncertainty. We also find a good correlation (significant at greater than the 99.5\% level) between the 30~GHz flux densities and the Fermi gamma-ray flux densities where sources are in common.

As expected, we find that the spectral index distribution for the sources broadens at higher frequencies. We ascribe this to the steepening of source spectral indices at different frequencies in different objects. We find 38 candidate Gigahertz-Peaked Spectrum sources, and also a significant number of inverted spectrum sources ($\sim 10$ per cent). It is clear that extrapolation from low frequency flux densities to higher frequencies assuming power law spectra cannot be relied upon to produce accurate results. This emphasises the need for high-frequency blind surveys to low flux densities. Such surveys are currently being carried out by the ATCA at 20~GHz in the southern hemisphere (see e.g. \citealp{2008Sadler}) and AMI at 15~GHz in the northern hemisphere \citep{2004Waldram,2009Waldram}, and will be carried out by the OCRA-F instrument \citep{2009Peela} and its successors at 30~GHz in the near future.

The flux densities of the sources within the sample described here will be useful for comparison \changed{with} point source measurements from the {\it Planck} satellite, which will make some of its most sensitive observations in the region covered here. The flat spectra of these sources should lead them to appear more \changed{frequently} in the Planck bands than other classes of sources. This prediction will be tested, and a comparison of the measurements can be carried out, when the {\it Planck} Early Release Compact Source Catalogue is released \changed{in January 2011.}\footnote{\changed{See \url{http://planck.ipac.caltech.edu/index.php?SiteSection=ERCSC}}}

\section*{Acknowledgements}
We are grateful for support of the OCRA project from a Royal Society International Project Grant. \changed{We are also grateful to the Polish Ministry of Science and Higher Education for their support of the OCRA project (grant number N N203 390434).} This research has made use of the NASA/IPAC Extragalactic Database (NED) which is operated by the Jet Propulsion Laboratory, California Institute of Technology, under contract with the National Aeronautics and Space Administration.

\onecolumn
\begin{longtable}[t]{l|c|c|r|r|r|r|l}

\caption[Flux densities for the CRATES sources]{Flux densities (in mJy) for the CRATES sources. See Section \ref{sec:crates_source_fluxes} for details.}
\label{tab:crates_fluxes}\\

Name & RA & Dec & $S_{1.4}$ & $S_{4.8}$ & $S_{8.4}$ & $S_{30}$ & Notes\\
\hline
& \multicolumn{2}{c}{J2000} & \multicolumn{4}{|c|}{Flux density/mJy} & \\
\hline
\endfirsthead

Name & RA & Dec & $S_{1.4}$ & $S_{4.8}$ & $S_{8.4}$ & $S_{30}$ & Notes\\
\hline
& \multicolumn{2}{c}{J2000} & \multicolumn{4}{|c|}{Flux density/mJy} & \\
\hline
\endhead

\multicolumn{8}{r}{{Continued...}}
\endfoot

\endlastfoot

\object{J0954+7435} & 09 54 47.5 & 74 35 57.1 & 1244.7 & 778 & 411.0 & 100.7$\pm$11.9 & c,g\\
\object{J1013+7427} & 10 13 30.7 & 74 27 26.9 & 125.8 & 87 & 63.4 & 106.2$\pm$11.7\\
\object{J1027+7428} & 10 27 24.1 & 74 28 26.1 & 190.1 & 211 & 143.5 & 72.0$\pm$9.8\\
\object{J1031+7441} & 10 31 22.0 & 74 41 58.3 & 213.2 & 250 & 103.9 & 123.0$\pm$12.3\\ 
\object{J1032+7218} & 10 32 31.9 & 72 18 18.9 & 121.4 & 100 & 71.0 & 62.6$\pm$7.1\\ 
\object{J1047+7238} & 10 47 47.5 & 72 38 13.0 & 36.7 & 108 & 31.2 & 58.3$\pm$6.1 & f\\ 
\object{J1048+7143} & 10 48 27.6 & 71 43 35.9 & 760.3 & 1900 & 1284.5 & 1148$\pm$100\phantom{.0} & w,f,c\\
\object{J1056+7401} & 10 56 18.7 & 74 01 00.2 & 75.7 & 67 & 58.1 & 23.8$\pm$5.2\\
\end{longtable}
\bsp
z
\newpage
\section*{Erratum}
In the paper `One Centimetre Receiver Array-prototype observations of the CRATES sources at 30~GHz' (MNRAS, 410, 2690), an error was made in the application of the gain-elevation corrections for data taken between August 2009 and June 2010. As this error affected both calibrator and source measurements, it has resulted in an offset of up to 1.75$\sigma$ in the mean flux densities for some sources (or up to 25 per cent of the mean source flux density for the faintest sources). It also resulted in an increased scatter between the measurements, which led to a small overestimate of the uncertainty on the mean flux densities. The average change of all of the source flux densities is a decrease of 1.2 per cent. This is sufficiently small that our scientific conclusions are unaltered.

\begin{figure}
\centering
\includegraphics[scale=0.23]{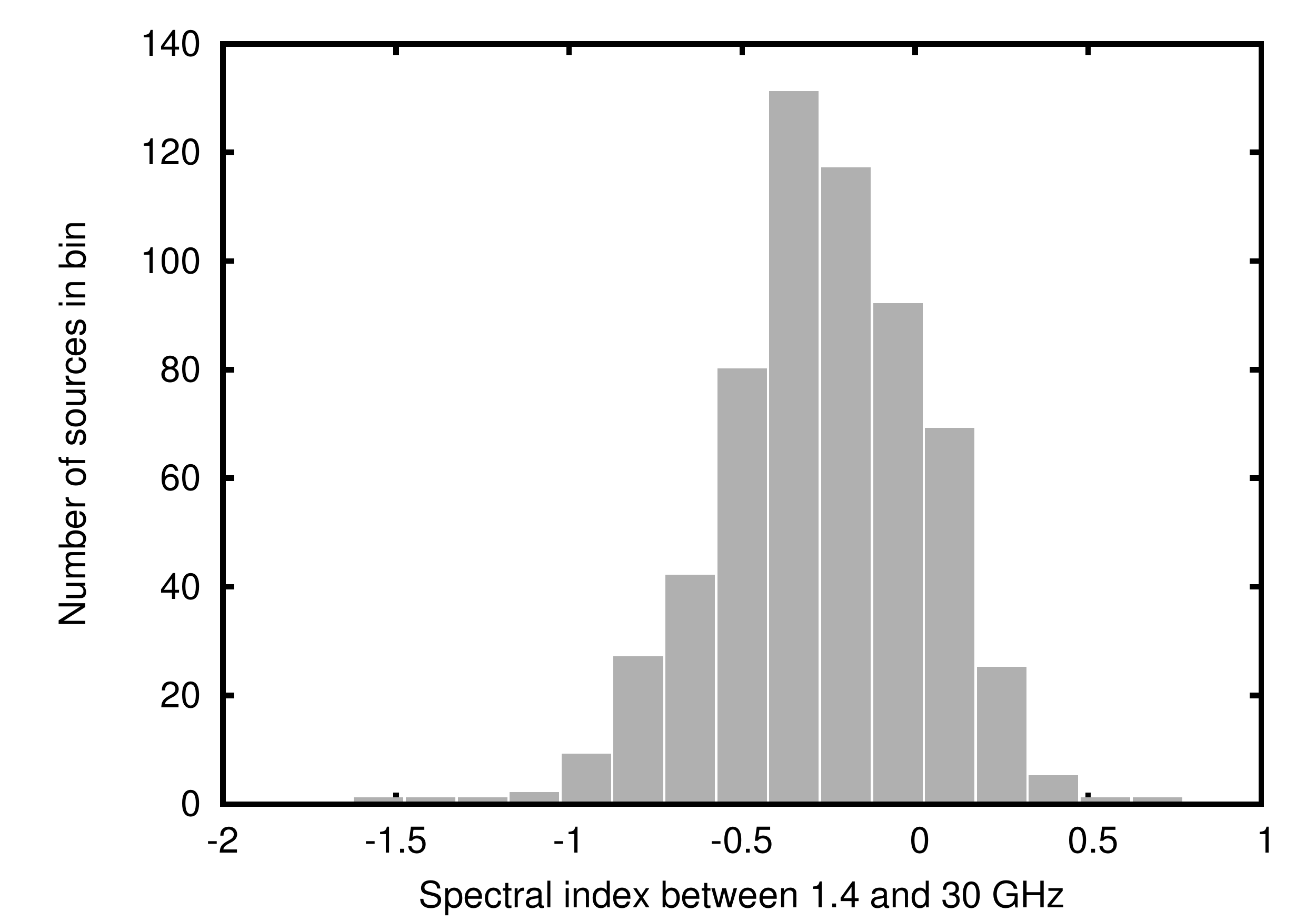}
\includegraphics[scale=0.23]{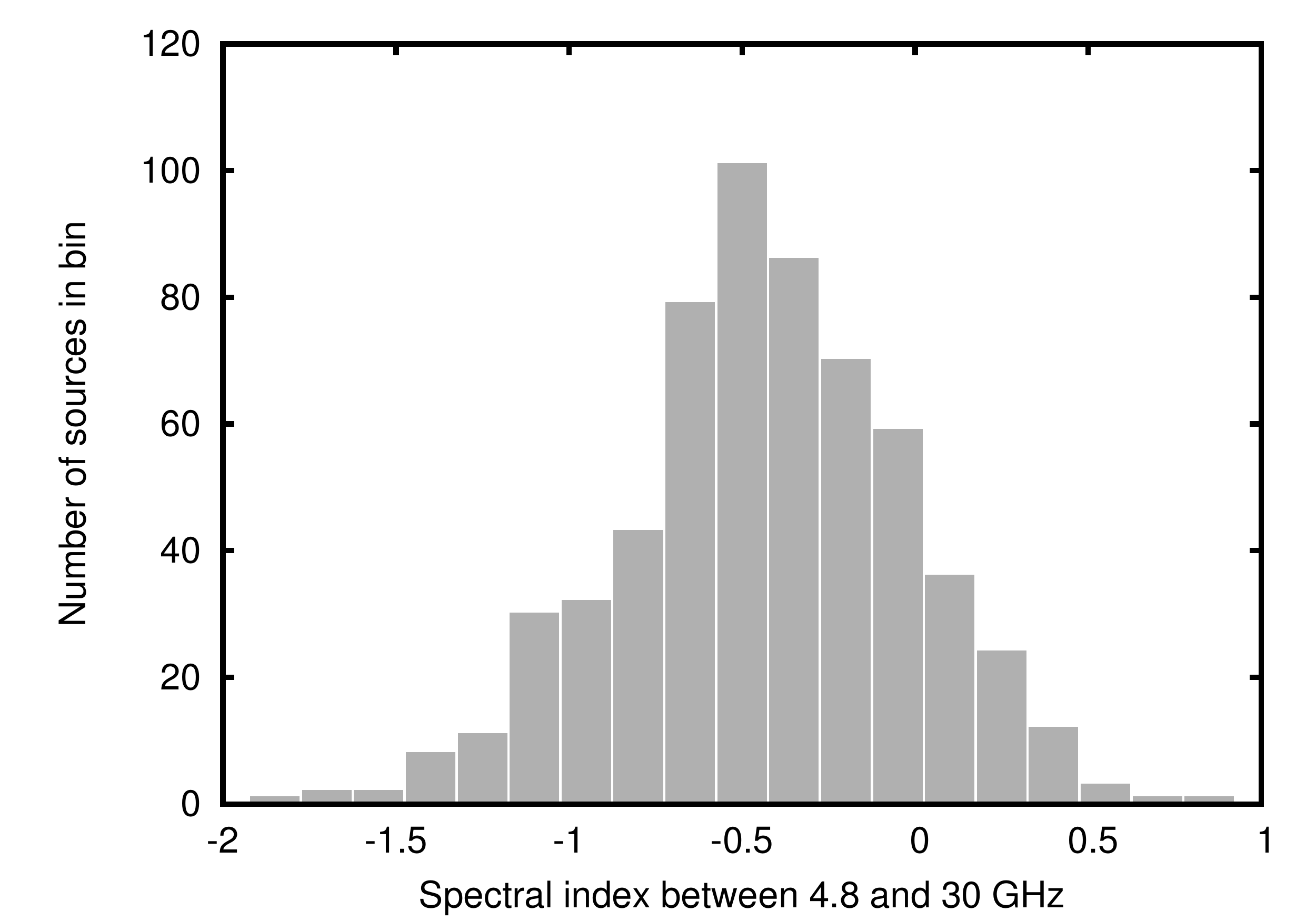}
\includegraphics[scale=0.23]{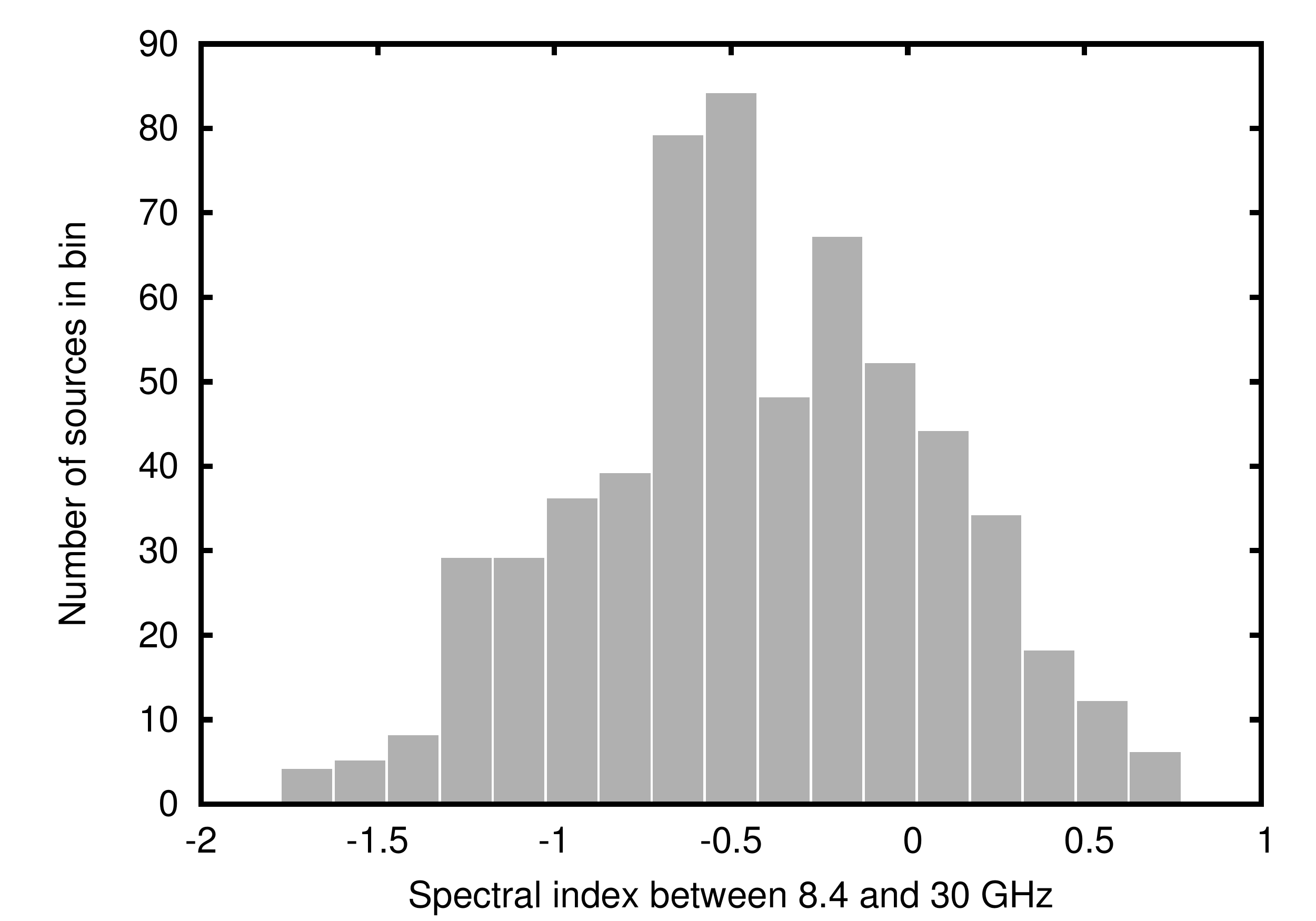}
\caption{Updated spectral index distributions for the sources in the CRATES sample observed with OCRA-p; the difference from Figure 5 in the original paper is negligible.}
\label{fig:crates_spectralindex_histogram_new}
\end{figure}

The updated spectral index distributions are shown in Figure \ref{fig:crates_spectralindex_histogram_new}, and the mean and spread for $\alpha_{1.4}^{30}$, $\alpha_{4.8}^{30}$, and $\alpha_{8.4}^{30}$ are unchanged from the values given in the paper. The other source  statistics change slightly, as follows:
 
\begin{enumerate}
\item 76 per cent of the sources have steepened at $\alpha_{8.4}^{30}$ compared with $\alpha_{1.4}^{4.8}$;
\item 33 sources (5.5 per cent) appear to be Gigahertz-peaked sources that peak between 4.8 and 8.4~GHz (defined by $\alpha_{1.4}^{4.8} > 0.5$ and $\alpha_{8.4}^{30} < -0.5$);
\item 63 sources (10 per cent) are flat or rising ($\alpha_{1.4}^{4.8} > 0$ and $\alpha_{8.4}^{30} > 0$);
\item 61 sources (10 per cent) are inverted ($\alpha_{1.4}^{4.8} < 0$ and $\alpha_{8.4}^{30} > 0$).
\end{enumerate}

The application of the Spearman-rank correlation test of the Fermi-radio correlation using the revised mean flux densities gives $\rho=0.69$ compared to the previous value of $\rho=0.71$. 

An updated version of Table 1 (comparing the OCRA measurements to those by WMAP) is included below. We have provided an updated version of the flux densities in Table 3 in the Supporting Information, as well as online at \url{http://www.jodrellbank.manchester.ac.uk/research/ocra/crates/}.

\begin{table}
\begin{tabular}{c|c|c|c|l}
{\bf Name} & {\bf OCRA} & {\bf WMAP K} & {\bf WMAP Ka} & {\bf Notes}\\
& {\bf 30~GHz} & {\bf 23~GHz} & {\bf 33~GHz} & \\
& \multicolumn{3}{|c|}{{\bf Flux density/Jy}} & \\
\hline
J1048+7143 & 1.14$\pm$0.09 & 1.21$\pm$0.61 & 1.11$\pm$0.73 & \\ 
J1101+7225 & 0.93$\pm$0.08 & 0.92$\pm$0.16 & 0.91$\pm$0.18 & a\\ 
J1302+5748 & 0.43$\pm$0.04 & 0.81$\pm$0.17 & 0.67$\pm$0.18 & a\\ 
J1343+6602 & 0.32$\pm$0.03 & 0.70$\pm$0.17 & 0.34$\pm$0.13 & Pa\\ 
J1344+6606 & 0.36$\pm$0.04 & as above & as above & \\
J1419+5423 & 0.62$\pm$0.12 & 0.86$\pm$0.18 & 0.89$\pm$0.25 & a\\ 
J1436+6336 & 1.04$\pm$0.08 & 0.68$\pm$0.23 & 0.50$\pm$0.28 & a\\ 
J1549+5038 & 0.57$\pm$0.05 & 0.85$\pm$0.14 & 0.64$\pm$0.11 & \\ 
J1604+5714 & 0.48$\pm$0.08 & 0.81$\pm$0.12 & 0.77$\pm$0.17 & a\\ 
J1625+4134 & 0.45$\pm$0.04 & 0.96$\pm$0.06 & 0.78$\pm$0.14 & \\ 
J1635+3808 & 2.47$\pm$0.28 & 3.72$\pm$0.86 & 4.06$\pm$1.08 & \\ 
J1637+4717 & 1.27$\pm$0.12 & 1.10$\pm$0.26 & 1.12$\pm$0.26 & \\ 
J1638+5720 & 1.66$\pm$0.15 & 1.67$\pm$0.43 & 1.68$\pm$0.48 &\\ 
J1642+3948 & 5.72$\pm$0.48 & 6.60$\pm$0.90 & 5.95$\pm$0.94 & \\ 
J1642+6856 & 4.78$\pm$0.41 & 1.82$\pm$0.68 & 1.85$\pm$0.85 & \\ 
J1648+4104 & 0.46$\pm$0.04 & 0.67$\pm$0.08 & 0.82$\pm$0.20 & a\\ 
J1653+3945 & 0.90$\pm$0.10 & 1.25$\pm$0.12 & 1.16$\pm$0.16 & a\\ 
J1657+5705 & 0.51$\pm$0.04 & 0.41$\pm$0.10 & -- &\\ 
J1658+4737 & 0.68$\pm$0.06 & 1.16$\pm$0.10 & 1.16$\pm$0.10 & \\ 
J1700+6830 & 0.39$\pm$0.05 & 0.32$\pm$0.07 & 0.51$\pm$0.08 & \\ 
J1701+3954 & 0.52$\pm$0.04 & 0.58$\pm$0.09 & 0.77$\pm$0.20 & \\ 
J1716+6836 & 0.60$\pm$0.05 & 0.58$\pm$0.11 & 0.51$\pm$0.16 & \\ 
J1727+4530 & 1.25$\pm$0.11 & 0.89$\pm$0.44 & 0.94$\pm$0.52 & \\ 
J1734+3857 & 1.05$\pm$0.09 & 1.20$\pm$0.11 & 1.26$\pm$0.24 & \\ 
J1735+3616 & 0.51$\pm$0.04 & 0.71$\pm$0.09 & 0.70$\pm$0.16 & \\ 
J1739+4737 & 0.67$\pm$0.05 & 0.86$\pm$0.12 & 0.77$\pm$0.20 & \\ 
J1740+5211 & 1.16$\pm$0.10 & 1.22$\pm$0.26 & 1.18$\pm$0.20 & \\ 
J1748+7005 & 0.47$\pm$0.04 & 0.50$\pm$0.15 & 0.63$\pm$0.16 & \\ 
J1753+4409 & 0.42$\pm$0.04 & 0.63$\pm$0.21 & 0.61$\pm$0.13 & a\\ 
J1800+3848 & 0.89$\pm$0.08 & 0.91$\pm$0.11 & 0.72$\pm$0.27 & a\\ 
J1801+4404 & 1.18$\pm$0.10 & 1.36$\pm$0.26 & 1.52$\pm$0.27 & \\ 
J1806+6949 & 1.52$\pm$0.13 & 1.42$\pm$0.14 & 1.36$\pm$0.22 & \\ 
J1812+5603 & 0.27$\pm$0.03 & 0.21$\pm$0.11 & 0.30$\pm$0.22 & \\ 
J1824+5651 & 1.60$\pm$0.14 & 1.47$\pm$0.26 & 1.24$\pm$0.18 & \\ 
J1835+3241 & 0.37$\pm$0.03 & 0.81$\pm$0.09 & 0.81$\pm$0.15 &\\ 
J1842+6809 & 0.95$\pm$0.09 & 1.29$\pm$0.27 & 1.40$\pm$0.30 & a\\ 
J1849+6705 & 3.68$\pm$0.30 & 1.63$\pm$0.66 & 1.81$\pm$0.78 & a\\ 
J1927+6117 & 0.58$\pm$0.05 & 0.95$\pm$0.26 & 0.94$\pm$0.22 & \\ 
J1927+7358 & 2.91$\pm$0.25 & 3.57$\pm$0.27 & 3.33$\pm$0.36 & \\ 
J2007+6607 & 0.63$\pm$0.06 & 0.78$\pm$0.09 & 0.57$\pm$0.08 & \\ 
J2009+7229 & 0.60$\pm$0.05 & 0.64$\pm$0.12 & 0.78$\pm$0.16 & \\ 
J2015+6554 & 0.36$\pm$0.03 & 0.75$\pm$0.14 & 0.79$\pm$0.15 & a\\ 
\end{tabular}
\caption{Comparison of 42 CRATES sources measured with OCRA-p that are also in the WMAP 7-year point source catalogue. The flux densities are given in Jy. Two sources -- J1343+6602 and J1344+6606 -- are combined in WMAP (denoted with ``P'').}
\label{tab:crates_wmap}
\end{table}

\label{lastpage}

\end{document}